# SoC Software Components Diagnosis Technology

Svetlana Chumachenko, Wajeb Gharibi, Anna Hahanova, Aleksey Sushanov
*Computer Engineering Faculty, Kharkov National University of Radioelectronics, Lenin Ave. 14, Kharkov, Ukraine, 61166, phone: (057) 70-21-421, (057) 70-21-326*
*E-mail: hahanov@kture.kharkov.ua; kiu@kture.kharkov.ua*

## Abstract

*A novel approach to evaluation of hardware and software testability, represented in the form of register transfer graph, is proposed. Instances of making of software graph models for their subsequent testing and diagnosis are shown.*

## 1. Introduction

There are technologies of hardware testing and testable design, which enable to solve the problem of SoC service effectively [1-10]. On the other hand, there are not effective models and methods of the given problem solving on the electronic technology market.

To realize testable design and diagnosis of SoC software components the universal model of software components representation in the form of register transfer and control graph is developed. An algorithm of software diagnosis is proposed. An instance of software diagnosis technology utilization is considered.

The research aim is adaptation of the hardware testing methods to the service of SoC software components.

The research problems: 1) Adaptation of Thatte-Abraham-Sharshunov register transfer model [4,5] to the solving of software testing problem; 2) Application of the model for faulty statements diagnosis on basis of use the fault detection table.

## 2. Software diagnosis technology

At development of large size software verification of development project on the correctness of statements is urgent problem. Complex software includes great many branches and verification of software on every logical path is rather complex problem. A method of faulty statements (errors or faults) searching for software that is based on representation of software algorithm in the form of graph structure for subsequent test generation and fault diagnosis is considered below on an example. Lets it is necessary to verify the software that realizes computation of the following sum of functions:

$$S = (x) + \omega(x),$$

$$x = \begin{cases} x + 3; & x < 2; \\ 2x - 3; & 2 \le x < 12; \\ -3x + 7; & x \ge 12; \end{cases}$$

$$\omega(x) = \begin{cases} \sin(x + \pi/3)\,, x < 2\pi/3; \\ \sin(\pi x) + 2\ , x \ge 2\pi/3. \end{cases}$$

One of the possible problem solution variants on C++ language is represented by the following listing:

*Listing 3.1.*

```
#include <iostream>
#include <math.h>
using namespace std;
int main()
{
    const double Pi=3.14159;
    double F, w, f, x;
    cin>>x;
if (x<2) f=x+3;
else if ((x>=2) && (x<12)) f=2*x-3;
else f=-3*x+7;
if (x<2./3.*Pi)
w=sin(x+Pi/3);
else w=sin(Pi*x)+2;
F=f+w;
cout<<F<<endl;
    return 0;
}
```

Lets an error takes place in a statement of computational part of software. Instead of the correct statement

```
else w=sin(Pi*x)+2;
```

the following one is written:

```
else w=sin(Pi*x) - 2;
```

It is necessary to detect faulty statement in program code by using the testing technology, based on the graph code model. Software diagnosis stages include 4 procedures below.

1. Making of register transfer graph.

Graph ribs are a set of code fragments or separate operations (Fig. 1); graph points are points of information monitoring (registers, variables, memory), which are used for forming of assertions too.

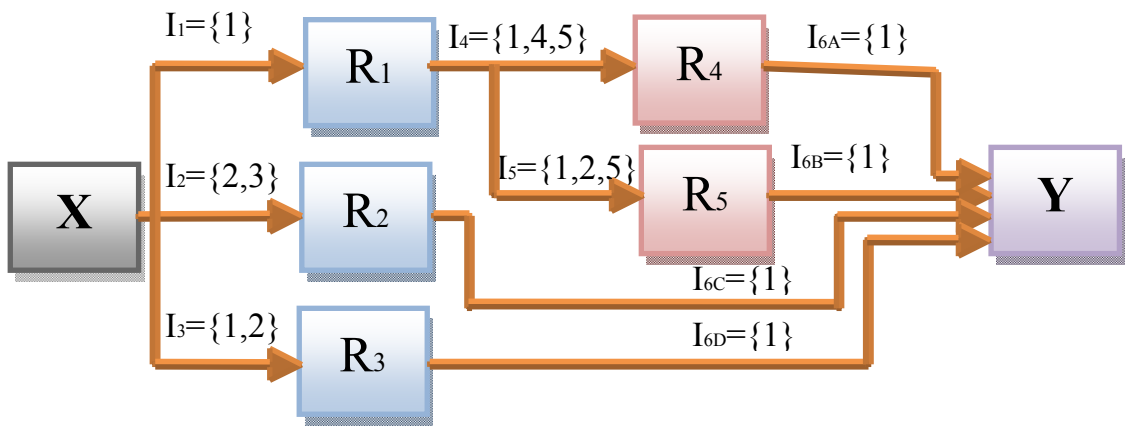

**Fig. 1. Register transfer graph**

A number of test points in the graph (registers, variables, memory) should be adequate to diagnose of given resolution. Otherwise it is necessary to carry out the analysis of register transfer graph testability for software and to determine the minimal additional quantity of observation lines for forming of assertions, which enable to detect faulty modules with given diagnosis resolution. Every rib (see Fig. 1) is marked by an arithmetic operation set: {1} – summation; {2} – multiplication; {3} – subtraction; {4} – division; {5} – obtainment of trigonometric sine. In a case when there is a branch in a program a number of outgoing ribs from a point is equal to quantity of adjacent sinks that is formed by branch statements in respective part of a program.

Thus, for the code fragment of the instance:

```
if (x<2) f=x+3;
else if ((x>=2) && (x<12)) f=2*x-3;
else f=-3*x+7;
```

there are three ribs, outgoing from the point X. Computational results $I_1, I_2, I_3$, which depend on the variable X, are checked in the points $R_1, R_2, R_3$ respectively. In a case of execution of the operation $I_1$ the following branch is realized:

```
if (x<2./3.*Pi) w=sin(x+Pi/3);
else w=sin(Pi*x)+2;
```

Then the general summation operation for all transactions is carried out regardless of which branch statements had been executed.

```
F=f+w;
```

The summation operation is executed on various ribs (the objects $I_{6A}, I_{6B}, I_{6C}, I_{6D}$), but all of them correspond to the same part of the program code. So, faultless execution an operation on a rib eliminates a fault on other three ones. On next stages of software diagnosis these objects are merged to $I_6$. The result are checked in the final point Y.

The method of software algorithm representation by graph structure enables to show all possible variants of software execution, as well as to simplify realization of next diagnosis stage of software and forming of minimal test.

2. Test synthesis and analysis. A set of ribs are written in the form of disjunctive normal form (DNF), where every term is one-dimensional path from input port to output, which covers a subset of internal lines: $P = X14Y \vee X15Y \vee X2Y \vee X3Y$. In the aggregate one-dimensional paths, represented in DNF, cover all possible transactions – graph points and ribs. An aggregate of code fragments or statements (activation instructions), written by disjunction, is brought to conformity with every rib. For instance, the path X14Y activates execution of operations on ribs $I_1, I_4, I_{6A}$. At that the ribs $I_1$ and $I_{6A}$ have only one statement, and consecutive execution of three statements corresponds to the identifier $I_4$. The test $P_1 = [(1)(1 \vee 4 \vee 5)(1)]$ that activates the path X14Y ensures the correctness check of all statements. Thus, the test of minimal covering of all graph points and ribs by commands, which activate graph ribs and therefore data movement to observation points, can be written:

$$P = [(1)(1 \vee 4 \vee 5)(1)] \vee [(1)(1 \vee 2 \vee 5)(1)] \vee$$
$$\vee [(2 \vee 3)(1)] \vee [(1 \vee 2)(1)].$$

Subsequent DNF transformation consists of removal of brackets to obtain complete test that enables to check transactions in a graph, which cover all points and ribs in various combinations:

$$P = (111 \vee 141 \vee 151) \vee (111 \vee 121 \vee 151) \vee$$
$$\vee (21 \vee 31) \vee (11 \vee 12).$$

The obtained test is redundant; it is not always acceptable for large size software, because of there is large quantity of test patterns. So, the ability to create minimal length test of given resolution is very important. Such test is formed by solving of the covering problem of all graph points and ribs and activation of code fragments sets. When testing it is supposed that hardware components, used in the software are faultless.

3. Fault detection table making. Fault detection table is oriented on verification of code fragments sets on ribs, which form data activation paths to the observation points (graph points). In compliance with comparison of experimental data of tested software and expected responses the output response vector V is formed. In a case of result failure on an observed line the respective coordinate of the vector V takes on a value “1” for the test pattern under consideration. The fault detection table of code fragments on complete test $P = X14Y \vee X15Y \vee X2Y \vee X3Y$, where test patterns are written in general form (a set of one-dimensional paths), is shown below:

| $T_i / I_j$ | $I_{11}$ | $I_{22}$ | $I_{23}$ | $I_{31}$ | $I_{32}$ | $I_{41}$ | $I_{44}$ | $I_{45}$ | $I_{51}$ | $I_{52}$ | $I_{55}$ | $I_{61}$ | V |
|---|---|---|---|---|---|---|---|---|---|---|---|---|---|
| X14Y | 1 | | | | | 1 | 1 | 1 | | | | 1 | 0 |
| X15Y | 1 | | | | | | | | 1 | 1 | 1 | 1 | 1 |
| X2Y | | 1 | 1 | | | | | | | | | 1 | 0 |
| X3Y | | | | 1 | 1 | | | | | | | 1 | 0 |
| Faults | | | | | | | | | 1 | 1 | 1 | | |

The symbolic notation $I_{jk}$ means execution of a statement that is on the rib $I_j$ and has index k. For instance, $I_{22}$ means execution of statement sequence of the rib $I_2$ at activation of the path X2Y and production operation that corresponds to the fragment of source program code:

```
else if ((x>=2) && (x<12)) f=2*x-3;
```

The diagnosis resolution for the test at the value of vector V = (0100) is determined by three possible faults: $F = I_{51} I_{52} I_{55}$. Value "1" of the vector V for a test-vector under consideration means that when issuing second pattern the activation of respective commands execution is took place. The minimal set of DNF terms, which make out all single faults of program fragments of a register transfer graph, is minimal diagnosis test. Next term set (here it coincide with complete test) makes out faults of all instructions, determined in DNF:

$P = (111 \vee 141 \vee 151) \vee (111 \vee 121 \vee 151) \vee (21 \vee 31) \vee (11 \vee 12)$.

Reduction impossibility is conditional on that removal any term does not provide activation of one or several fragments. Then complete and extended fault detection table is made that is formed by a term set above. Every obtained test pattern is divided on parts – terms. First test pattern $(111 \vee 141 \vee 151)$ consists of three terms: (111), (141) and (151). Every of them has own position in a column. All possible executable operations, which are designated $I_{ik}$, where j – rib identifier in a graph, k – statement that transforms data on j-th rib, is distinguished across. The graph path to which a term under consideration is applied is considered. For instance, term (141) is applied to first test pattern that activates the path X14Y. The extended fault detection table is:

| $T_i \setminus I_i$ | $I_{11}$ | $I_{22}$ | $I_{23}$ | $I_{31}$ | $I_{32}$ | $I_{41}$ | $I_{44}$ | $I_{45}$ | $I_{51}$ | $I_{52}$ | $I_{55}$ | $I_{61}$ | V |
|---|---|---|---|---|---|---|---|---|---|---|---|---|---|
| $111_1$ | 1 | | | | | 1 | | | | | | 1 | 0 |
| 141 | 1 | | | | | | 1 | | | | | 1 | 0 |
| $151_1$ | 1 | | | | | | | 1 | | | | 1 | 0 |
| $111_2$ | 1 | | | | | | | | 1 | | | 1 | 1 |
| 121 | 1 | | | | | | | | | 1 | | 1 | 1 |
| $151_2$ | 1 | | | | | | | | | | 1 | 1 | 1 |
| $21_1$ | | 1 | | | | | | | | | | 1 | 0 |
| 31 | | | 1 | | | | | | | | | 1 | 0 |
| 11 | | | | 1 | | | | | | | | 1 | 0 |
| $21_2$ | | | | | 1 | | | | | | | 1 | 0 |

Every term number means execution of a statement on respective graph rib. First nimber "1" provides activation of the statement {1} $I_1$, so opposite respective column "1" is put. Column values of the extended fault detection table are moved from the FDT of code fragments that is defined on complete generalized test. But coordinate value is written for every test term. Extended fault detection table enable to show the results of every test pattern execution and to simplify the fault detection procedure with given resolution.

4. Diagnosis. In compliance with numbers of "1' in the output response vector V quantity of disjunctive CNF terms is formed. Every term is line-by-line writing of faults by logical operation "OR", which influence on distortion of output functional signals. Then transformation CNF to DNF by the Boolean algebra is carried out:

$$F = (I_{11} \vee I_{51} \vee I_{61})(I_{11} \vee I_{52} \vee I_{61})(I_{11} \vee I_{55} \vee I_{61}) =$$
$$I_{11} \vee I_{11}I_{55} \vee I_{11}I_{61} \vee I_{11}I_{52} \vee I_{11}I_{52}I_{55} \vee I_{11}I_{52}I_{61} \vee$$
$$\vee I_{61}I_{11} \vee \vee I_{11}I_{61}I_{51} \vee I_{11}I_{61} \vee I_{51}I_{11} \vee I_{11}I_{51}I_{55} \vee$$
$$\vee I_{11}I_{51}I_{61} \vee I_{11}I_{51}I_{52} \vee \vee I_{51}I_{52}I_{55} \vee I_{51}I_{52}I_{61} \vee$$
$$\vee I_{51}I_{61}I_{11} \vee I_{55}I_{61} \vee I_{51}I_{61} \vee I_{11}I_{61} \vee I_{11}I_{55}I_{61} \vee$$
$$\vee I_{11}I_{61} \vee I_{11}I_{52}I_{61} \vee I_{52}I_{55}I_{61} \vee I_{52}I_{61} \vee I_{61}I_{11} \vee$$
$$\vee I_{61}I_{51} \vee I_{61}.$$

To reduce the obtained set of possible faults the Boolean algebra laws are used:

$A \wedge A = A;\ A \vee B = B \vee A;\ (A \vee B)C = AC \vee BC;$
$(A \vee B) \vee C = A \vee (B \vee C);\ A \vee A = A;$
$(A \wedge B) \vee A = A;\ (A \vee B) \wedge A = A$, it enables to obtain the expression:

$$F = I_{11} \vee I_{11}I_{55} \vee I_{11}I_{61} \vee I_{11}I_{52} \vee I_{11}I_{52}I_{55} \vee$$
$$\vee I_{11}I_{52}I_{61} \vee I_{61}I_{11} \vee I_{11}I_{61}I_{51} \vee I_{11}I_{61} \vee$$
$$\vee I_{51}I_{11} \vee I_{11}I_{51}I_{55} \vee I_{11}I_{51}I_{61} \vee I_{11}I_{51}I_{52} \vee$$
$$\vee I_{51}I_{52}I_{55} \vee I_{51}I_{52}I_{61} \vee I_{51}I_{61}I_{11} \vee I_{55}I_{61} \vee$$
$$\vee I_{51}I_{61} \vee I_{11}I_{61} \vee I_{11}I_{55}I_{61} \vee I_{11}I_{61} \vee I_{11}I_{52}I_{61} \vee$$
$$\vee I_{52}I_{55}I_{61} \vee I_{52}I_{61} \vee I_{61}I_{11} \vee I_{61}I_{51} \vee I_{61} =$$
$$= I_{11} \vee I_{51}I_{52}I_{55} \vee I_{61}.$$

Then such elements $I_{jk}$ from F, which are executed in other test patterns with value $V_i = 1$, are removed. A set of objects, contained the operations, which transform data at program execution uniquely and correctly, is formed:

$$H = \{X14Y, X2Y, X3Y\} = \{(141) \vee (151) \vee (21_1) \vee (31) \vee$$
$$\vee (11) \vee (21_2)\} = I_{11} \vee I_{22} \vee I_{23} \vee I_{31} \vee I_{32} \vee I_{44} \vee I_{45} \vee I_{61}.$$

After the reduction a single DNF term is obtained:

$$F' = F \setminus H = (I_{11} \vee I_{51}I_{52}I_{55} \vee I_{61}) \setminus (I_{11} \vee I_{22} \vee I_{23} \vee I_{31} \vee I_{32} \vee I_{44} \vee I_{45} \vee I_{61}) = I_{51}I_{52}I_{55}.$$

It means that the software functions with error at execution one of the statements {1,2,5} on the rib $I_5$.

Really, an error takes place on linear program part that is applied to a rib of the statement sequence $I_5$, namely $I_{51}$ – execution of subtraction instead of summation.

More exact diagnosis (to within statement) is possible if to use the greater quantity of test points that complicates diagnosis because of necessity to make longer tests. The proposed method enables to analyze software on presence of errors in the code and helps to detect their location. Testing and verification of software is the main problem at programming, and its solving enables to raise software quality and to obviate unforeseen results of its execution. The proposed method is based on representation of software algorithm by the graph structure, where ribs are statement sequences or code fragments, and points are information monitoring points for making of assertions. Creation of minimal quantity of test patterns enables to decrease time of fault detection. At that tests have to cover all possible transactions. Test points quantity has to be minimal and sufficient for diagnosis of given resolution.

## 3. Conclusion

The innovative technologies of software testable design, based on effective test development and verification of digital system-on-a-chip components, are considered.

1. The universal model of software and hardware component in the form of directed register transfer and control graph, on which the testable design, test synthesis and analysis problems can be solved, is represented.
2. The technology of software testing and diagnosis on basis of synthesis the graph register transfer models is proposed.
3. The practical importance of proposed methods and models is high interest of the software companies in innovative solutions of the effective software testing and verification problems above.